# Load balancing mechanisms in fog computing: A systematic review

Mostafa Haghi Kashani, Ahmad Ahmadzadeh, and Ebrahim Mahdipour*

**Abstract**—Recently, fog computing has been introduced as a modern distributed paradigm and complement to cloud computing to provide services. Fog system extends storing and computing to the edge of the network, which can solve the problem about service computing of the delay-sensitive applications remarkably besides enabling the location awareness and mobility support. Load balancing is an important aspect of fog networks that avoids a situation with some under-loaded or overloaded fog nodes. Quality of Service (QoS) parameters such as resource utilization, throughput, cost, response time, performance, and energy consumption can be improved with load balancing. In recent years, some researches in load balancing techniques in fog networks have been carried out, but there is no systematic review to consolidate these studies. This article reviews the load-balancing mechanisms systematically in fog computing in four classifications, including approximate, exact, fundamental, and hybrid methods (published between 2013 and August 2020). Also, this article investigates load balancing metrics with all advantages and disadvantages related to chosen load balancing mechanisms in fog networks. The evaluation techniques and tools applied for each reviewed study are explored as well. Additionally, the essential open challenges and future trends of these mechanisms are discussed.

———————————— ◆ ————————————

## 1 INTRODUCTION

FOG computing, which extends from the cloud and is a geographically distributed paradigm, brings networking power and computing into the network edge, closer to end-users and IoT devices both because of being supported by wide-spread fog nodes [1]. Most of the data, in cloud-only architectures, requiring processing, analysis, and storage, are transmitted to the cloud servers, which may have an influence on latency, security, mobility, and reliability adversely. With the existence of location-aware and delay-sensitive applications, the cloud on its own comes across some problems to meet the extremely-low latency requirements of these applications; the proximity of the fog layer to the Internet of Things (IoT) devices may remarkably decrease latency and meet the needs of extremely-low [2], [3]. Fog computing always interacts with and supports the cloud, creating a novel generation of applications and services.

Nowadays, in fog computing environments, users need applications that give quick responses whenever they want to access anything and work fast. To improve QoS factors in a fog network significantly, we can use an efficient load balancing strategy because load balancing is regarded as an important issue. Many studies have been done to balance the cloud computing load because the load on the cloud increases enormously [4]. Being heterogeneous and dynamic, the fog networks cannot directly apply most of the load balancing mechanisms of cloud computing; The goal of load balancing in fog environment is to distribute the coming load between available fog nodes or cloud, based on one mechanism, to avoid overload or under-load of fog nodes. This mechanism can maximize throughput, performance, and resource utilization while minimizing response time, cost, and energy consumption.

We have had no Systematic Literature Review (SLR) of research on load balancing mechanisms in fog computing that makes it hard to evaluate identifying the gaps of studies, different trends, and specifically future dimensions of load balancing in fog environment. In addition, regarding the ever-increasing need for load balancing in fog computing, we are required to investigate a research agenda for load balancing mechanisms in fog computing. An SLR can identify, categorize, and synthesize a comparative review of state-of-the-art studies. It also makes knowledge transfer possible in the research community [5], [6]. This SLR is conducted with the aim of *identification, taxonomic classification, and systematic comparison of the existing researches that focus on planning, executing, and validating the fog systems load balancing*. We especially aim at answering the questions below by conducting a methodological overview of the existing studies:

- What are the main practical motivations for load balancing in fog computing?
- Which kind of classification in research approaches can be applied in fog systems load balancing?
- What evaluation metrics are applied in load balancing mechanisms of fog computing?
- What are popular evaluation tools applied in load balancing mechanisms in fog computing?
- What measurement techniques are used to assess the load balancing in fog computing?
- What are the open issues, future trends, and challenges

———————————————

- *Mostafa Haghi Kashani, Ahmad Ahmadzadeh, and Ebrahim Mahdipour are with the Department of Computer Engineering, Science and Research Branch, Islamic Azad University, Tehran, Iran. E-mail: mh.kashani@srbiau.ac.ir, ahmet.ahmadzade@gmail.com, mahdipour@srbiau.ac.ir*
- **The corresponding author is Ebrahim Mahdipour.*

of load balancing in fog computing?

Guidelines in [5], [7], [8] were followed. Our purpose is to have a systematic identification and taxonomic classification of the evidence available on load balancing in fog computing and to have a holistic comparison to analyze the limitations and potential of the existing study. It makes an SLR of the current study by concentrating on the suggested techniques, solutions, and methods in load balancing in fog computing. Therefore, **36** studies are chosen, categorized, and compared by applying a characterization taxonomy. The characterization taxonomy is composed of four groups, including approximate, exact, fundamental, and hybrid, which are derived and refined by following a qualitative evaluation of the included researches, some famous references [9], and our current experience with previous systematic studies [10], [11], [12]. The research synthesis led to a knowledge base of recent research mechanisms, techniques, methods, experiences, and best practices that were applied in load balancing in fog computing. Furthermore, open challenges and future trends related to load balancing methods in fog systems are discussed. The results related to this systematic study are beneficial for

- Scholars in fog/cloud systems, who require the identification of relevant researches. Presenting a research systematically procures a corpus of knowledge that is necessary for developing theories and solutions, analyzing research implications, and establishing future directions.
- Practitioners who are eager for understanding the available techniques and methods with tool support and their limitations in supporting load balancing mechanisms to fog-based environments.

The structure of the study is organized as follows: Section 2 shows the background of fog computing, load balancing, and metrics definition. Also surveys related to this study are presented in this section. Section 3 illustrates the methodology of the research. Also, Section 4 depicts the selected load balancing methods in fog computing in four categories. Section 5 refers to the results and comparisons of techniques, and then, in Section 6, open issues are outlined. Further, in Section 7, limitations of review are explained. At the end, Section 8 shows the conclusions of this study.

## 2 BACKGROUND AND RELATED WORK

The concept and structure of fog computing (Section 2.1) and load balancing (Section 2.2) are discussed and explained in this part. In addition, we link the existing review studies for load balancing in fog computing (Section 2.3) based on a systematic exploration (Section 2.4).

### 2.1 Fog Computing

Due to the unprecedented amount of data and the connection of over 50 billion devices to the Internet (based on Cisco estimation), handling that much of data with traditional computing models, like cloud computing, distributed computing, etc. is difficult [13]. Often privacy gaps, high communication delay, related network traffic loads that connect cloud computing to end-users for unpredictable reasons with the recent expansion of services related to IoT (like smart cities, eHealth, industrial scenarios, smart transportation systems, etc. [14]) are some challenges that affect cloud computing performance. To refer to some of cloud computing limitations and to bring cloud service traits so much closer to "Things", as it is referred to, including cars, mobile phones, embedded systems, sensors, etc., the research community has suggested the fog computing concept [1].

Fog computing is regarded as a platform bringing cloud computing to end-users' vicinity. "Fog", as a term, has an analogy with real-life fog and was initially introduced by Cisco [1]. When the fog is nearer to the earth, clouds are up above in the sky and, interestingly, fog computing applies this concept, when the virtual fog platform is located closer to end users just between end-users' devices and the cloud. In a similar definition, fog computing is suggested to make computing possible at the network edge, to send new services and applications specifically for the Internet future [15]. Bonomi, et al. [1], to give a more appropriate definition of fog computing for the first time, said that fog computing was not exclusively located at the network edge. However, it was a virtualized platform providing networking services, storage, and computations among the data centers and end devices of conventional cloud computing.

*Fog computing is most often mistaken for edge computing*, but we have major differences between the two. Fog computing applications are run in a multi-layer architecture that disconnects and meshes the software and hardware functions, permitting the dynamic reconfigurations for diverse applications while executing transmission services and intelligent computing. Edge computing, on the other hand, creates a direct transmission service and manages special applications in a fixed logic location. While Fog computing is hierarchical, edge computing is limited to a few peripheral devices. Besides networking and computation, fog computing deals with the control, storage, and acceleration of data-processing [16], [17]. An IoT client or smart end-device, to recognize fog computing from other computing standards, needs to utilize the following characteristics but not all of them while consuming a fog computing service [13], [18].

- *Low latency and contextual location awareness*: Awareness of fog node latency costs for communicating with other nodes and their logical location within the entire system context makes fog computing offer the lowest possible latency. The necessity of low latency in applications is the key in fog computing that supports and points with wealthy services at the network edge. Analysis and response to information created by fog nodes are really speedier than that of a centralized data center or cloud service since they are mostly co-located with the smart user-devices.
- *Geographical distribution*: In contrast with a more centralized cloud, fog computing gives services to geographically-identifiable, distributed deployments. In other words, the fog computing will have a key part in conveying high-quality spilling services to motioned vehicles, via access points and proxies geographically situated along tracks and highways.
- *Heterogeneity*: Fog computing backs up processing and collection of diverse information form components obtained via various kinds of network communication capabilities.
- *Federation and interoperability*: Services must be unified across domains, and components of fog computing have to be capable of interoperation since supporting certain services requires the cooperation of different providers.
- *Real-time interactions*: Real-time interactions in fog computing applications are involved instead of batch processing.
- *Fog-node clusters, agility, and scalability of federated*: Fog computing is naturally adaptive, at the level of cluster or cluster-of-clusters, network condition varieties, data-load changes, resource pooling, and supporting elastic compute, to mention but a few of supported adaptive functions.
- *Predominance of wireless access*: Fog system is exceptionally located in wireless IoT access networks. Even though fog computing is utilized in wired environments, a massive number of IoT wireless sensors need distributed compute and analytic.
- *Support for mobility*: Supporting mobility techniques, like locator/ID separation protocol, decoupling the identity of the host from the identity of location, needs a system that has a distributed directory system, which is necessary for numerous applications in fog computing for direct communication with mobile devices.

### 2.1.1 Fog Node

As a layered model, fog computing enables all-out access to a wide shared range of scalable computing resources. This model helps the arrangement of distributed, delay-aware services and applications, and is composed of fog nodes (virtual or physical), located between centralized (cloud) services and smart end-devices. The context-aware fog nodes back up a regular communication and data management system. These nodes would be sorted in clusters - either horizontally (in order to back federation), vertically (in order to back isolation), or relatively to fog node delay-distance to the smart user-devices. The fog node is the key element in fog architecture [15]. Fog nodes can be either *virtual elements* like virtual machines (VMs), virtualized switches, or *physical elements* such as switches, gateways, closely matched with access networks or intelligent end-devices, and able to make computing resources to the devices mentioned. For the sake of facility in deploying the capability of fog system, fog nodes are required to back one or some of the features below [10], [18]:

- *Autonomy*: Fog nodes are capable of independent operations to make local decisions, at the level of node or cluster-of-nodes.
- *Heterogeneity*: Fog nodes, with various form factors, can be used in a vast range of environments.
- *Hierarchical clustering*: Fog nodes, cooperating as a continuum, back hierarchical structures, with various layers creating diverse subsets of service functions.
- *Manageability*: Complicated systems manage and orchestrate fog nodes, and these systems can automatically do the most common operations.
- *Programmability*: At different levels, fog nodes are programmable by different stakeholders -like end-users, network operators, and domain experts.

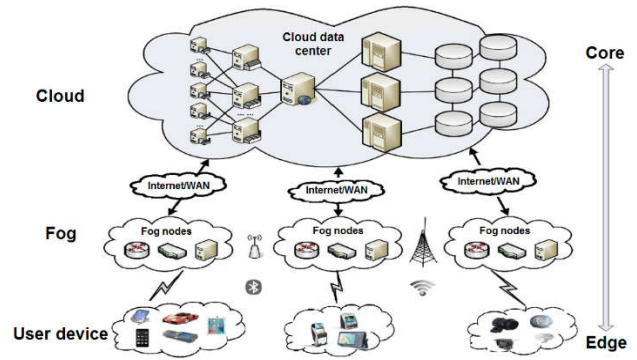

Fig. 1. The architecture of fog network [10], [13]

### 2.1.2 Fog Architecture

Fog computing architecture reference model is an important study topic. Recently, a wide range of architectures has been suggested for fog computing, mostly obtained from a structure with three layers [11], [16]. Fog network expands cloud services to the network edge by suggesting a fog layer between cloud and user devices. As it can be observed, Fig. 1 illustrates the fog architecture hierarchically [16], having three layers as follows:

- *Cloud layer*: The layer of cloud computing is composed of different storage devices and high-performing servers and creates several services of applications. It bears robust storage and computing abilities to back the permanent storage of a large amount of information and extended computation analysis. However, it should be noted that all computing and storage tasks do not pass the cloud that is not the same as traditional cloud computing architecture [19].
- *Fog layer*: The fog layer is located at the network edge, which consists of a couple of fog nodes like access points, routers, switches, gateways, etc. They are spread between cloud and end devices. In order to get services, end devices might easily connect with fog nodes. They are capable of computing, storing, and transmitting the received sensed data. The latency-sensitive applications and real-time analysis can be performed in a fog layer. In addition, we can refer to the connection between the cloud data center and the fog nodes by the IP core network. In order to get more robust storage and computing capabilities, fog nodes have the responsibility of interacting and cooperating with the cloud.
- *User device*: The layer of a user device is so close to the physical environment and end-user. This layer is composed of different IoT devices, like, sensors, cellphones, smart automobiles, cards, and readers. Although cellphones and smart vehicles have got computing capabilities, they are utilized as just smart sensing devices. Generally and geographically, they are widely distributed and responsible for sensing feature data related to events or physical objects and for transferring them to upper layers to be processed and stored.

Here in the architecture, all end devices or smart objects are connected with fog nodes by technologies with wired or wireless connection access such as 3G, 4G, wireless LAN, ZigBee, Bluetooth, and Wi-Fi. Wireless or wired communication technologies to help the interconnection and intercommunications of fog nodes. IP core network helps fog nodes each to be linked with the cloud [10].

## 2.2 Load Balancing

In fog system, load balancing facilitates the distribution of workload on resources equally, aiming to provide services continually if the service component fails, and it is done by provisioning and de-provisioning instances of applications along with proper resource utilization. Because data centers procure diversities between hosts and show special features of traffic, an appropriate mechanism of load balancing is needed in fog computing to refine the performance of applications and utilization of the network [20]. To evade any overload or under-load on resources, load balancing, as a mechanism, spread the workload onto different resources. Load balancing, which distributes the load among different resources, is implemented either in physical equipment or software [21].

The load balancing has some goals, including throughput maximization, response time minimization, and traffic optimization. Consumption optimization in the server-side resources, request processing time minimization, and scalability improvement in the distributed environment are some other purposes of the technique of load balancing [22]. In fog networks, load balancing may have various methods that can be of static or dynamic nature or both. In static methods, with primary information about the system as a necessary feature, the rule should be programmed in the load balancer because the user's behavior is not predictable, and methods of static load balancing are not necessarily efficient in the network. Further, the dynamic methods outperform the static methods because of the dynamic distribution of load based on the pattern that is programmed in the load balancer [23]. Mechanisms of dynamic load-balancing apply current system state to this end, and they use especial policies including [24]:

- *Transfer*: It defines the conditions based on which a task has to be sent from one node to the other. The arriving tasks that enter the transfer policy are transferred or processed based on a determining rule that relies on each node workload. The policy deals with task migration and rescheduling.
- *Selection*: It defines whether a task should be sent or not and also regards a couple of elements to select a task, like the amount of overhead needed for migration, time of task execution, and total nonlocal system calls.
- *Location*: It defines under-loaded nodes and then sends tasks to these nodes. In aimed nodes, the availability of essential services for task rescheduling or migration is checked.
- *Information*: The complete information, considering system nodes, is collected in this policy and is used by other policies to make a decision. This policy determines the time at which the information should be collected. Various policies have some relationships that are mentioned below:

Transfer policy grabs incoming tasks and determines whether to transfer them to a remote node to balance load or not. If not eligible to be transferred, the task will be locally processed. When the transfer policy decides to transfer a task, the location policy would be triggered to locate a remote node to process the task. The task is locally processed when a far partner cannot be found, or the task is transferred to a remote node. The necessary information is provided by information policy for location and transfers both to help them to make a decision.

### 2.2.1 Load Balancing Metrics

Several metrics are needed to assess a mechanism of load balancing and weigh it in comparison with mechanisms before to show which mechanism is better and to recognize the pros and cons related. The metrics need some qualitative paradigms. Various qualitative metrics are used in articles, like response time, cost, energy, etc. The essential metrics for load balancing in fog computing are stated below:

- *Response time*: This issue is described by the interval starting from the acceptance of a request (or task) to the response to a request for a server in fog environment [25].
- *Cost*: The payment of money to ask for an action that is required to do [26].
- *Energy consumption*: It refers to the energy consumption amount in a fog network. Energy consumption can be decreased by an effective load balancing mechanism [27].
- *Scalability*: It shows how the system is capable of accomplishing a load balancing mechanism with a couple of hosts or machines [28].
- *Security*: It is the quality side of service that procures non-repudiation and confidentiality via authentication involving parties and message encryption [29].
- *Flexibility*: Fog nodes that always connect to a system pro tempore that incline to leave periodically. So, to reflect both nodes that are joined newly and nodes revocation, this mechanism has to be flexible [30].
- *Resource utilization*: It represents the maximum utilization of the resources available in a cloud system [10].
- *Deadline*: The latest time when a service request in the fog system can be completed [10].
- *Processing time*: The duration in which a service request in the fog system is executed entirely [14].
- *Reliability*: The ability of a fog network to perform its required requests in a defined time and a specified condition [31].
- *Throughput*: We can refer to the maximum requested service rate might be processed in the fog system as throughput [14].
- *Availability*: A rise in resource application or service requests can maintain the system performance that shows the capability of a computing system [14].

TABLE 1
RELATED REVIEWS OF LOAD BALANCING IN FOG/EDGE COMPUTING

| Review type | Ref. | Main idea | Publication year | Article selection process | Reviewed articles | Open issue | Covered year |
|---|---|---|---|---|---|---|---|
| Survey | [32] | Fog | 2019 | Not clear | 9 | Not presented | Not-mentioned |
|  | [33] | Fog | 2019 | Not clear | 10 | Not presented | Not-mentioned |
|  | [34] | Edge | 2020 | Not clear | 29 | presented | Not-mentioned |
| SLR | Our study | Fog | - | Clear | 36 | Presented | 2013-August 2020 |

## 2.3 Review Studies for Load Balancing Methods in Fog/Edge Computing

No systematic review was found on load balancing mechanisms in fog computing. Thus, the decision was made to investigate the review studies existing on load balancing mechanisms in fog/edge computing to conduct this systematic review. The studies were summarized in Table 1 as review studies based on surveys [32], [33], [34].

Chandak and Ray [32] presented a survey of load balancing techniques in fog computing. They also introduced some of the evaluation parameters and simulation tools used for load balancing methods in the fog system. In addition, Baburao, et al. [33] surveyed some of the techniques of service migration, load balancing, and load optimization in fog computing. Furthermore, Pydi and Iyer [34] reviewed load balancing methods in edge systems, including security-based, traffic load-based, optimization-based, heuristic-based, joint load-based, multi-access-based, allocation-based, dynamic load-based, and distributed-based techniques. They introduced some of the evaluation factors and tools. Also, the authors discussed some of the future trends and research gaps.

Table 1 presents a summary of studied articles that depicted some parameters like the type of reviews, main ideas, year of publications, the process of article selection, reviewed articles, open issues, and covered years of every study. No single article yet has created based on systematic method in the field of fog. So, this review study, for the first time, look through the load balancing approaches in the fog by applying the systematic method.

## 2.4 The Motivations for a Systematic Study on Load Balancing Methods in Fog Computing

Needing a systematic review leads to *identification*, *classification*, and *comparison* of the existing evidence on the load balancing mechanisms in fog environment. It concentrates on the classification and comparison of load balancing approaches in the fog system. In order to show that a resembling review has not been yet reported, we surfed the Google Scholar (on 1/9/2020) with the search strings below:

(*fog* <OR> *edge*) [**AND**]
(*load* <OR> *balancing*) [**AND**]
(*survey* <OR> *review* <OR> *overview* <OR> *challenges* <OR> *trends* <OR> *issues* <OR> *study*)

Among the obtained review studies, no one was related to any of our research questions in Section 3. Regarding the significance of load balancing mechanisms in fog networks, the consolidation of the existing evidence on load balancing mechanisms in the fog system is necessary.

## 3 RESEARCH METHOD

Contrary to the procedure in a non-structured review, a systematic review [5] decrease the partiality and follows an exact sequence of methodological stages to research literature. A systematic review depends on truly-defined and assessed review protocols for extracting, analyzing, and documenting the results, as shown in Fig. 2. We obtained the guidelines in [5], [35] with a three-stage study procedure, including *planning, conducting, and documenting*. The study is accomplished by an external assessment of the results of each phase. A clear classification of the reviewed studies is provided, which is a foundation for a comparative analysis of researches based on the dimensions of our analysis that are also subject to external assessment. The planning and conducting stages of the methodology are summarized to perform this systematic review (as shown in Fig. 2). In terms of data summary, the results are illustrated in Section 5, and for findings and research implications the results are depicted in Section 6. Based on the proposed taxonomy, the data are collected and synthesized, as explained in Section 4.

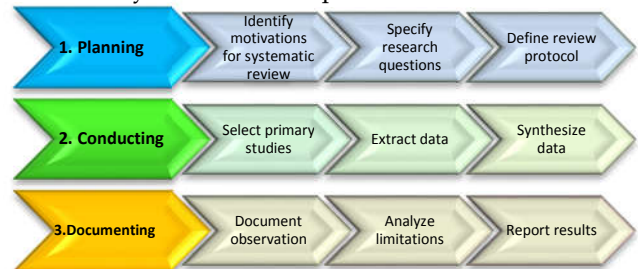

Fig. 2. Our research method steps

### 3.1 Planning the Review

Planning begins with knowing the motivations for a systematic study and the results in a review protocol as defined below:

#### 3.1.1 Identify the Motivations for the Systematic Review.

In Section 2.4, we identify the motivation and justify the contribution of this systematic review.

#### 3.1.2 Specifying the Research Questions

The research questions (RQs) define our motivation, i.e., answers give us an evidence-based review of load balancing mechanisms. Six research questions are defined that clarify the basis for obtaining the strategy of the search for extracting literature, as shown in Table 2. The aim of investigating each question is outlined by motivation. A comparative analysis, on the other hand, permits an analysis of the collective influence of the research, which is presented in terms of comparison features.

TABLE 2
RESEARCH QUESTIONS

| |
|---|
| *RQ 1*-What are the main practical motivations for load balancing in fog computing? |
| *RQ 2*-Which kind of classification in research approaches can be applied in fog systems load balancing? |
| *RQ 3*-What evaluation metrics are applied in load balancing mechanisms of fog computing? |
| *RQ 4*-What popular evaluation tools are applied in load balancing mechanisms in fog computing? |
| *RQ 5*-What measurement techniques are used to assess the load balancing in fog computing? |
| *RQ 6*-What are open issues, future trends, and challenges of load balancing in fog computing? |

### 3.1.3 Define Review Protocol

Based on the purposes, the RQs and the study scope were specified to make the search strings for extracting literature. A protocol was also developed for a systematic study by following [5] and our experience with systematic reviews [10], [11], [12]. As proposed in [7], [35], the protocol was externally assessed before execution. An external expert was asked for feedback, who was experienced in conducting systematic reviews in a field that overlapped with fog computing. The feedback given is reflected in a defined protocol. A pilot study of the systematic survey was performed, containing 20 percent of the included researches. The purpose of this pilot study was primarily reducing the partiality among the researchers and improving the characterization method for data collection. The study scope was expanded, the search methods were improved, and the exclusion/inclusion criteria were refined during the experimental studies.

## 3.2 Conducting the Review

The second phase is conducting that starts with articles selection and leads to data extraction and information synthesis:

### 3.2.1 Select Primary Studies

Conducting, as research methodology second phase, begins with selecting articles and leads to data extraction. This subsection aim is presenting the procedures of searching and choosing articles in the second phase of the systematic review. To select the articles, a two-step guideline is followed:

- *Initial selection.* The search strings that follow are found among academic databases to locate articles having these strings in their abstracts, titles, and keywords. Accordingly, famous online academic databases like ACM, IEEE, ScienceDirect, Springer, Google Scholar, Taylor & Francis, and Wiley are used. 915 articles were extracted primarily (see Table 3). We also regarded the online-published articles from 2013 to August 2020. 2013 was chosen because fog computing was introduced in 2012 [1].

| *fog [**AND**]* |
|---|
| *(load <OR> balancing <OR> balanced <OR> balancer)* |

- *Final selection.* We examined 915 articles extracted from the previous step and applied the inclusion/exclusion criteria (as mentioned in Table 4). Next, we investigated the articles fully, applied quality assessment; only articles that had mentioned the assessment details and techniques explicitly were chosen; therefore, **36** related studies were finally selected to be evaluated qualitatively.

TABLE 3
SEARCH RESULTS ON DIGITAL LIBRARIES

| No | Academic database | Result |
|---|---|---|
| 1 | ACM | 20 |
| 2 | IEEE | 177 |
| 3 | ScienceDirect | 134 |
| 4 | Springer | 37 |
| 5 | Google Scholar | 481 |
| 6 | Taylor & Francis | 3 |
| 7 | Wiley | 63 |
| | *Total* | *915* |

TABLE 4
INCLUSION/EXCLUSION CRITERIA

| | |
|---|---|
| Inclusion | • Research articles that present techniques or innovative solutions on load balancing mechanisms in fog computing<br>• Peer-reviewed articles in conferences and JCR-indexed journals<br>• Articles published between 2013 and August 2020 |
| Exclusion | • Review articles, editorial articles, short articles (less than six pages), white articles, and non-English articles<br>• Research articles that do not mention solutions and methods to improve load balancing in fog computing explicitly<br>• Books, book chapters, and theses |

### 3.2.2 Data Extraction and Synthesis

As mentioned in Section 4, a structured format was designed and followed [5] based on characterization dimensions to record the obtained data from chosen articles. By investigating the limitations and potentials of current studies and the reflections on future studies, an organized comparative analysis was formed, which provided us with an investigation of the collective research impact.

## 4 CLASSIFICATION OF THE SELECTED STUDIES

A structured classification of the related literature is defined here. Because of the diverse literature on load balancing mechanisms of fog computing, it is not easy to structure the related works systematically. Based on the available literature, Fig. 3 shows the framework of the suggested classification scheme. Four main categories were recognized: approximate, exact, fundamental, and hybrid methods. It was a natural move to review the literature from these perspectives since most studies in this domain deal with the issues from each perspective that permits categorizing the reviewed articles under common superordinate. However, other taxonomies are also possible (e. g. *centralized, semi-distributed,* and *distributed,* or *system state* and *who initiated the process*). This section reviews 36 selected articles based on the criteria mentioned before, and their main features, differences, evaluation parameters, tools, pros and cons, and evaluation techniques are defined.

## 4.1 Approximate Methods

In this part, studies related to approximate methods, including stochastic, probabilistic, and statistic techniques are performed. In Sections 4.1.1 and 4.1.2, we investigate stochastic methods, including heuristic and meta-heuristic related to the research field, respectively. Then, in Section 4.1.3, probabilistic/statistic methods are re-viewed.

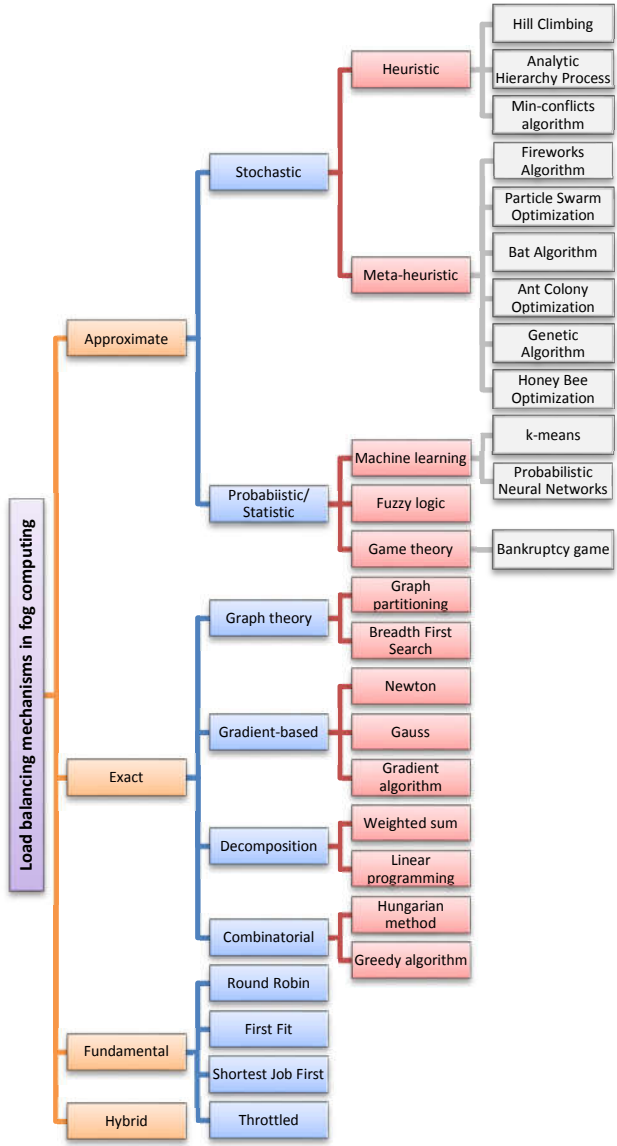

Fig. 3. Classification of load balancing mechanisms in fog computing

*4.1.1 Heuristic Methods*

Heuristic methods are totally made by "experience" for special optimization problems, intending to find the best solution to the problem through "trial-and-error" in an optimal amount of time [36]. The solutions in heuristic approaches might not be the best or optimal solution; however, they can be much better than an educated guess. Heuristic approaches make use of the problem particularities. As exact approaches consume a huge amount of time to get the optimal solution, heuristic approaches are preferable, gaining near-optimal solutions in an optimal amount of time [36]. Some of the heuristic method in the literature reviewed includes Hill Climbing [37] Min-conflicts [38], and Analytic Hierarchy Process [39]. In this section, selected heuristic-based methods are discussed.

For three-layered architecture, Zahid, et al. [37] proposed a framework that was composed of a distributed fog layer, a centralized cloud layer, and a consumer layer.

A hill-climbing load-balancing method, decreasing the processing time (PT) and response time (RT) of fogs to consumers, was suggested here. There was a tradeoff, though, in RT and microgrid cost. So, this article aims to procure the request load balancing on fog nodes. In addition, Kamal, et al. [38] presented a load balancing scheduling mechanism as Min-conflicts scheduling. This mechanism makes use of a heuristic method solving a constraint satisfaction problem. The architecture proposed is composed of cloud, fog, and end-users as three layers.

In order to reduce the delay of data streams from IoT devices to the applications, Banaie, et al. [39] proposed multiple vacation-based queuing systems to model the performance of a fog system. To speed up user's access to sensor data, they leveraged multi-gateways architecture along with a resource caching policy in the IoT domain. A load-balancing scheme based on the analytic hierarchy process (AHP) method was also employed to provide global load fairness among the network entities. Further, Oueis, et al. [40] focused on improving users' quality of experience by referring to load balancing in a fog environment. They considered multiple users that require computation offloading, in which the whole requests have to be processed by local computation cluster resources. Also, Xu, et al. [41] suggested a virtual machine (VM) scheduling mechanism to balance the load in the cloud-fog system.

*4.1.2 Meta-heuristic Methods*

A meta-heuristic method, as a higher-level heuristic method, problem-independent and can be applied to a wide range of problems. Today's "Meta-heuristics" indicates all modern higher-level methods [36]. We have two major parts in modern meta-heuristics: diversification and intensification [42]. It is important to have a balance between diversification and intensification to gain an influential and effective meta-heuristic method. A meta-heuristic method investigates the whole solution space; a different set of solutions should be produced, and search has to be heightened near the neighborhood of the optimal or near-optimal solutions. Some of the meta-heuristic methods in the literature reviewed include Particle Swarm Optimization [43], [44], Fireworks Algorithm [45], and Bat Algorithm [46].

He, et al. [43] integrated fog and software-defined network (SDN) to tackle the challenges. For the sufficient use of the SDN and cloud/fog architecture on the Internet of vehicle, they presented an SDN-based modified constrained optimization particle swarm optimization method. Also, based on fog network, Wan, et al. [44] suggested an energy-aware load balancing and scheduling approach. Firstly, an energy consumption model was proposed on the fog node that was related to the workload, and then an optimization function was formulated to balance the manufacturing cluster load. Then, the improved particle swarm optimization (PSO) method was applied to gain a good solution, and to achieve tasks, the priority has to be given to the manufacturing cluster.

Shi, et al. [45] integrated fog and SDN to the cloud-based mobile face recognition architecture for solving the delay problem. They also formulated the load balancing in the SDN and fog/cloud system as an optimization problem and proposed using a fireworks algorithm (FWA) based on SDN centralized control for solving the load balancing problem. Furthermore, Yang [46] proposed a three-layered architecture based on fog/cloud network and big medical data, including cloud, fog, medical devices. In the proposed architecture, the bat algorithm used the load balancing strategy to perform the initial setup of bat population data, which improved the quality of the solution in the initial sample.

TABLE 5
APPROXIMATE LOAD BALANCING METHODS IN FOG COMPUTING AND THEIR PROPERTIES
In the evaluation column, P=>Prototype, S=>Simulation, and N=>Not-mentioned.

| Method | | Article | Main idea | Evaluation | Tool | Advantage | Disadvantage |
|---|---|---|---|---|---|---|---|
| Stochastic | Heuristic | [37] | Hill climbing load balancing algorithm based on fog system | S | CloudAnalyst | • Low response time<br>• Low processing time | • Low scalability<br>• Low security |
| | | [38] | Load balancing in heuristic Min-conflicts optimizing method | S | CloudAnalyst | • Low response time<br>• Low cost<br>• Low latency | • High complexity<br>• Requires experience and knowledge |
| | | [39] | A load-balancing scheme based on the AHP method for Multiple Gateways in a fog network | S | MATLAB | • Low response time<br>• Low energy consumption | • High complexity<br>• Low availability |
| | | [40] | Resource management based on load distribution for fog clustering | S | Not-mentioned | • Customizable design<br>• Low energy<br>• Low complexity<br>• Low latency | • Inability to recover from database corruption |
| | | [41] | A heuristic VM scheduling mechanism to provide load balancing | S | CloudSim | • Avoid bottleneck<br>• High resource utilization<br>• Avoid overload | • Not supporting the positive and negative impacts of VM migration<br>• Imbalances of the positive and negative impacts in service migration |
| | Meta-heuristic | [43] | Load balancing mechanism based on SDN in cloud/fog network | S | Not-mentioned | • Low response time<br>• High mobility<br>• Improve the QoS<br>• Low latency | • Low security<br>• Low scalability |
| | | [44] | Scheduling and load balancing for fog-based smart factory | P | Work robots | • Optimal scheduling<br>• Low energy | • The efficiency of broadcast mode is low |
| | | [45] | Load balancing based on SDN in fog/cloud system | S | MATLAB | • Low response time<br>• Low cost<br>• Low latency | • Low performance |
| | | [46] | A fog/cloud system and big medical data based on bat algorithm considering load balancing | S | MATLAB | • Low latency | • High complexity<br>• The possibility of a bottleneck<br>• Low scalability<br>• Low reliability |
| Probabilistic/Statistic | Machine learning | [30] | Load balancing method in large-scale fog environment | S | iFogSim | • High scalability<br>• Low response time<br>• High flexibility<br>• High resource utilization | • Low performance<br>• High overhead |
| | Fuzzy logic | [47] | A load balancer based on fuzzy logic in fog computing | N | Not-mentioned | • Low energy consumption<br>• Low latency | • Low reliability<br>• Low security |
| | Game theory | [48] | Load balancing in fog network for great machine-type communications | S | Not-mentioned | • Low response time<br>• Low energy<br>• Low execution time | • High complexity |

*4.1.3 Probabilistic/statistic methods*

In this section, load balancing mechanisms based on probabilistic/statistic methods, including machine learning [30], fuzzy logic [47], and game theory [48] are discussed.

Li, et al. [30] examined the fog infrastructure runtime features and proposed a self-similarity-based load balancing (SSLB) technique for large-scale fog systems. They proposed an adaptive threshold policy and a corresponding scheduling method to guarantee SSLB efficiency successfully. Further, Singh, et al. [47] introduced a load bal-ancer based on fuzzy logic using different levels of tuning and designing of fuzzy controls in fog networks. The proposed fuzzy logic model was used to conduct link analysis as interconnects for managing traffic. Abedin, et al. [48] formulated a fog load balancing problem, to minimize load balancing cost of fog environment empowered by narrow-band Internet of thing (NB-IoT). Firstly, the time resource scheduling problem in NB-IoT was modeled as a Bankruptcy game. Subsequently, the transportation problem was solved by applying Vogel's approximation technique that locates a feasible load balancing solution to guarantee optimal assignment of jobs in the fog environment.

The classification of the above-mentioned articles and essential factors, in analyzing the approximate load balancing mechanisms in fog computing, are depicted in Table 5.

**4.2 Exact Methods**

Exact methods can optimally solve optimization problems. Each optimization problem might be solved by ap-plying the exact search, but the bigger the size of the instances, the more time it takes forbiddingly to get the optimal solution. The exhaustive search is considerably slower than the exact methods [49]. Some of the exact methods in literature reviewed include graph theory [50], [51], gradient-based [52], [53], [54], decomposition [55], [56], [57], [58], and combinatorial [59], [60]. In this section, the studied articles that are based on exact methods are summarized below:

The graph partitioning theory was used by Ningning, et al. [50] to make the fog computing load balancing method on the basis of dynamic graph partitioning. The authors showed that, after cloud atomization, the fog computing framework could flexibly build the system network, and the dynamic load balancing mechanism is capable of configuring the system and reducing node migration caused by system changes. In addition, a load balancing technique was suggested by Puthal, et al. [51] to validate the edge data centers (EDCs) and find less loaded EDCs for the allocation of tasks. This technique is more useful, compared to other techniques, in locating less loaded EDCs for the allocation of tasks. It not only increases the efficacy of load balancing but also improves security via destination EDCs authentication.

Moreover, a workload balancing model was presented by Fan and Ansari [52] in a fog network for minimizing the data flow latency in processing procedures and communications through the association of IoT devices to appropriate base stations. Barros, et al. [53] used fog computing as a means of reducing the logical distance between consumption spot and central distribution. IoT devices in the network edge have higher efficacy and lower cost to manage the power flow information. They evaluated the Gauss-Seidel and Newton-Raphson method performance aiming to develop computations in real-time of the load flow problem with the assistance of fog. Beraldi and Alnuweiri [54] studied load balancing among fog nodes and addressed the special challenges resulting from the fog system. Particularly, they applied randomized-based load balancing protocols that leveraged the power-of-random choice property. Based on parallel exploring, they proposed sequential probing contrasting the classical randomization protocols.

Chen and Kuehn [55] considered the downlink of the cache-enabled fog-radio access network (F-RAN) and investigated minimizing the consumption of power to communicate green. Based on channel states, an efficient load balancing algorithm was suggested. With the proposed algorithm, the increase in cache memory for greater content hitting rate was considered an economical method in achieving greener networks. Further, Maswood, et al. [56] presented a Mixed-Integer Linear Programming (MILP) model in the fog/cloud environment to improve the bandwidth cost in routing, link utilization of network, and server resource utilization. They considered load balancing strategy jointly at the server level and the network both.

Also, Sthapit, et al. [57] proposed solutions for some situations in which the cloud or fog is unavailable. Firstly, by using a network of queues, the sensor network was modeled, then, considering the load balancing, a linear programming technique was applied to make scheduling decisions. Chen, et al. [58], by using connected car systems as an evocative application, showed that mobility patterns of vehicles could be utilized for performing periodic load balancing in fog servers. A task model was presented by them for solving the scheduling problem at the server level, not at the device level.

Also, Dao, et al. [59] presented an adaptive resource balancing (ARB) model to maximize serviceability in F-RANs in which the resource block (RB) utilization within remote radio heads (RRHs), by applying the Hungarian method and backpressure technique, are balanced, considering a time-varying network topology issued by potential RRH mobility. Further, a load-balancing scheme was proposed by Mukherjee, et al. [60] to define the tradeoff between computing delays and transmission in F-RANs.

Table 6 describes the classification of the aforementioned studies and the influential elements for analyzing the exact load balancing methods in fog computing.

## 4.3 Fundamental Methods

In the existing literature, some researches on load balancing strategy in fog computing is based on simple methods without complex computations that are classified in the fundamental methods. They include such methods as Shortest Job First [61], Throttled, Round Robin (RR), and First Fit [62], [63]. In this part, the selected fundamental methods are reviewed.

Nazar, et al. [61] suggested a load balancing algorithm modified the shortest job first (MSJF) to manage user's request load between VMs at fog level aiming to optimize the suggested cloud and fog performance based on integrated architecture. In addition, Ahmad, et al. [62] proposed an integrated cloud and fog based platform to manage energy effectively in the smart buildings. The first fit (FF) method was applied for load balancing that chooses VMs based on partitioning memory blocks. In the cloud/fog-based model, smart buildings having many apartments consist of IoT devices that were regarded. Also, a decentralized scheduling architecture was presented by Chekired, et al. [64] for energy management of electric vehicles (EVs) based on the fog system paradigm, where, by applying priority-queuing model, an optimal load balancing algorithm (LBA) was performed. Neto, et al. [65] proposed a Multi-tenant Load Distribution approach for Fog networks (MtLDF) regarding specific multi-tenancy needs like latency and priority. Further, they presented case studies to illustrate the proposed method applicability compared to a latency-driven load distribution mechanism.

Furthermore, an approach was presented by Batista, et al. [66] based on executing load balancing need for the fog of things (FoT) platforms by programmability for distributed IoT environments by applying SDN. The authors addressed the problems by applying FoT load balancing mechanism and assessed response time and lost samples as two metrics.

TABLE 6
EXACT LOAD BALANCING METHODS IN THE FOG SYSTEM AND THEIR PROPERTIES
In the evaluation column, F=>Formal verification and S=>Simulation.

| | Method | Article | Main idea | Evaluation | Tool | Advantage | Disadvantage |
|---|---|---|---|---|---|---|---|
| Graph theory | Graph partitioning | [50] | Dynamic load balancing technique in the fog network | S | JMeter | • Low cost<br>• High flexibility | • High complexity<br>• Extra overhead at execution time |
| | Breadth First Search | [51] | Load balancing technique in edge data centers and task allocation in fog environment | S, F | Scyther, MATLAB | • High security<br>• Low latency | • Not proposing lightweight security solutions and not improving load balancing performance |
| Gradient_based | Gradient algorithm | [52] | Workload balancing scheme in fog/IoT model | S | Not-mentioned | • Low response time<br>• Low energy | • Bottleneck<br>• Low scalability<br>• Low availability |
| | Newton-Raphson, Gauss-Seidel | [53] | Dynamic Load Flow in fog-based SGs | S | MATLAB | • High reliability<br>• Low latency<br>• Low cost | • High complexity<br>• High execution time |
| | Fixed point algorithm | [54] | Load balancing method based on Sequential Randomization in fog nodes | S, F | Custom simulator with Python | • Low cost | • Low reliability |
| Decomposition | Weighted sum | [55] | Load balancing and radio unit selection in the downlink of F-RAN | S | Not-mentioned | • Low energy<br>• Low cost | • deactivation should be considered for lower operational energy consumption<br>• Increasing the cache memory for higher content hitting rate |
| | | [56] | A MILP model to improve resource utilization in a fog network considering load balancing | S | AMPL/CPLEX | • High Resource utilization<br>• Low cost | • Low scalability<br>• Low availability |
| | Linear programming | [57] | Load balancing for computation in an edge system | S | NS-3 | • High performance<br>• Low energy | • The performance boost also comes at a similar energy cost |
| | | [58] | Load balancing for minimizing total runtime and deadline in fog-based vehicle systems | S | Not-mentioned | • Low energy<br>• Low response time<br>• Low latency | • Low reliability<br>• Priorities cannot be set |
| Combinatorial | Hungarian Method | [59] | Resource balancing scheme in F-RAN | S | Not-mentioned | • High reliability<br>• Low response time<br>• High throughput | • Low performance<br>• In different network conditions, the amount of service migration is not considered |
| | Greedy algorithm | [60] | Load balancing mechanism for F-RAN | S | Not-mentioned | • Low response time<br>• Low energy | • High execution time<br>• Low reliability |

Tariq, et al. [63] designed a fog based environment to cover a vast area of six regions in the world, each regarded as a single region with many consumers that sends requests on fog to access to the needed resources. A load-balancing method was proposed to select VMs efficiently inside a fog system for the consumers to receive a fast response with minimum latency. As well as for fog-cloud based architecture, Verma, et al. [67] suggested an efficient load balancing (ELB) technique. It used the technique of information replication to maintain those data in fog networks that reduced the total dependency on massive data centers.

Table 7 describes the classification of the aforementioned studies and the influential elements for analyzing the fundamental load balancing methods in fog computing.

## 4.4 Hybrid Methods

To accomplish load balancing in fog networks, hybrid methods apply such various methods as approximate, exact, and fundamental [68], [69], [70], [71], [72], [73]. Studies with hybrid methods are reviewed in this section.

Naqvi, et al. [68] introduced fog computing to increase cloud computing processing speed that is a cloud computing complement. Fog nodes, each having four to nine VMs, use the service broker policies for request processing. The use of the Ant Colony Optimization (ACO) load balancing method, throttle, and RR balances VMs load. Also, Abbasi, et al. [69] concentrated on fog computing application to a smart grid (SG) consisting of a distributed generation environment recognized as a microgrid. The aim of this study was to improve delay time, response time, and resource utilization. The proposed VM load balancing technique performed better than the other techniques mentioned in the results. Ali, et al. [70] proposed a four-layered SG-based architecture to improve communication between consumers and electricity company, and this model covers a huge area of residents. Three load balancing mechanisms were applied for allocation of VM, and the service broker policies applied for simulations are dynamically reconfigurable and were the closest to data centers. For resource allocation Zubair, et al. [71] used Genetic Algorithm (GA), throttle, and RR for load balancing mechanism by applying bin pack techniques. In this study, an SG was integrated with fog, and the cloud-based model and three places with some buildings were considered.

Moreover, based on the Q-learning and GA, Talaat, et al. [72] presented a load balancing scheme using the dynamic resource allocation approach in a fog-based healthcare environment. The load balancing scheme continuously monitors network traffic, collects information about each server load, and controls incoming requests and distributes them among the existing servers applying a dynamic resource allocation method.

TABLE 7
FUNDAMENTAL LOAD BALANCING METHODS IN THE FOG SYSTEM AND THEIR PROPERTIES
In the evaluation column, E=>Example application and S=>Simulation.

| Method | Article | Main idea | Evaluation | Tool | Advantage | Disadvantage |
|---|---|---|---|---|---|---|
| MSJF | [61] | The shortest job first-based method for providing load balancing | S | CloudAnalyst | • Low response time<br>• Low cost | • Low performance<br>• Longer processes have a more waiting time |
| Throttled,<br>RR,<br>First Fit | [62] | Resource allocation in fog/cloud system considering load balancing | S | CloudSim | • Low response time<br>• Low energy | • High cost |
| LBA algorithm | [64] | Load balancing method in distributed fog architecture | S | NS-2<br>MATLAB | • High scalability<br>• Low energy<br>• Low response time<br>• Low latency | • Not stable at evening peak hours<br>• Standard case without scheduling is not stable at peak hours |
| MtLDF algorithm | [65] | A load-balancing method for fog Environment | E | Java platform | • Low energy<br>• Low latency<br>• Low cost | • Not improve the load balancing across the Fog-Cloud layers<br>• Not consider some features such as disk I/O operations<br>• Low resource utilization |
| FoT Load Balancing algorithm | [66] | Load balancing method in FoT System | S | Mininet | • Low response time<br>• High performance in IoT | • Low scalability<br>• Not developed a solution for IoT scenarios<br>• The solution was not compared with traditional load-balancing algorithms<br>• Low scalability |
| RR,<br>Throttled | [63] | Load balancing based on priority in fog/cloud systems | S | CloudAnalyst | • Low response time<br>• Low latency<br>• Low energy<br>• High flexibility | • Results compared to throttled are not much better |
| ELB algorithm | [67] | Effective data replication considering load balancing | S | CloudSim | • Low cost<br>• Low response time<br>• Low processing time | • Low security<br>• Low privacy<br>• Low reliability |

TABLE 8
HYBRID LOAD BALANCING METHODS IN THE FOG SYSTEM AND THEIR PROPERTIES
In the evaluation column, S=>Simulation

| Method | Article | Main idea | Evaluation | Tool | Advantage | Disadvantage |
|---|---|---|---|---|---|---|
| RR,<br>Throttle,<br>ACO | [68] | Metaheuristic method in the fog-cloud system for load balancing | S | Java platform | • Low cost<br>• Low response time | • High complexity<br>• Low security |
| Throttled,<br>RR,<br>Particle Swarm Optimization | [69] | Load balancing method for load stabilization in fog environment | S | CloudAnalyst | • Low energy<br>• Low response time | • The limitation of this article is that the results of the presented broker policy do not outperform the results of existing broker policies |
| RR,<br>Honey Bee Optimization | [70] | State-based load balancing method for SG energy management in fog | S | CloudAnalyst | • High performance<br>• Low cost<br>• Low response time<br>• Low latency | • There are some changes of values in different fogs, but overall the average cost remains the same |
| RR,<br>Throttled,<br>Hybrid GA | [71] | Integration of fog-cloud based system for load balancing by applying bin packing and genetic methods | S | CloudAnalyst | • Low response time<br>• Low cost<br>• High security<br>• Low latency | • High complexity<br>• Difficult to show branching and looping |
| Weighted RR,<br>Q-learning,<br>GA | [72] | A load balancing strategy based on genetic algorithm and Q-learning in the fog-based healthcare system | S | MATLAB | • High Resource utilization<br>• Low response time | • High complexity<br>• Low scalability |
| Fuzzy logic,<br>Probabilistic Neural Networks | [73] | Load balancing method in real-time fog system based on neural network and Fuzzy logic algorithms | S | iFogSim | • Low latency<br>• Low cost<br>• Low response time<br>• Low energy | • QoS can be measured in terms of reduced IoT service delay<br>• Can make it more distributed |

In addition, Talaat, et al. [73] presented an influential load balancing strategy (ELBS) for fog system appropriate for applications in healthcare. ELBS attempted to attain important load balancing in fog environment through caching algorithms and real-time scheduling. The authors presented ELBS for fog environment appropriate for applications of the healthcare system.

Table 8 shows the classifications of the above-mentioned studies and the influential elements in analyzing the hybrid methods of load balancing in the fog networks.

## 5 ANALYSIS OF RESULTS

This section analyzes the results of the systematic review. We give a review of the selected articles in Section 5.1, and then introduce a comparison of the articles in Section 5.2 by answering the RQs.

## 5.1 Overview of the Selected Studies

The following complementary questions (CQs) are considered to examine the state of research on load balancing in fog network:

- *CQ1*: What is the status of research on load balancing in fog networks?
- *CQ2*: How was the distribution of studies per publication channel in load balancing in fog computing?
- *CQ3*: When did research on load balancing in fog networks become active in the computing community?

### 5.1.1 Temporal Overview of Studies

As illustrated in Fig. 4, where 58% of the articles are associated with IEEE, 39% to Springer, and 3% to ScienceDirect. Figure 5 illustrates the overtime classification of the articles in every category, including IEEE, Springer, and ScienceDirect, published between 2013 up to August 2020. The work on load balancing in fog network started with solutions with load balancing for small cell fog/cloud computing [40] in 2015. It should be noted that just 1 article was published in 2015, 4 articles in 2016, 3 articles in 2017, 21 articles in 2018, 2 articles in 2019, and 5 articles in 2020.

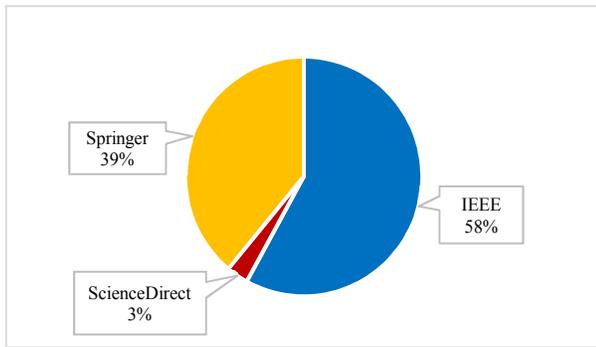

Fig. 4. The percentage of articles per publisher

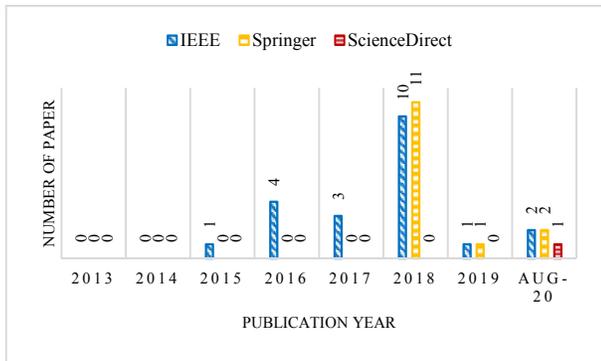

Fig. 5. The distribution of articles over time-based on publisher

### 5.1.2 Publication Fora

Most of the studies on load balancing in fog computing were published in IEEE Access, BWCCS, ChinaComm, JAIHC, GLOBECOM, INCoS, 3PGCIC, NBiS, and other journals and conferences (as mentioned in Table 9). Among 36 studied articles, 15 articles were published in JCR-indexed journals including IEEE Access (impact factor: 3.745), ChinaComm (impact factor: 2.024), TII (impact factor: 9.112), IoT-J (impact factor: 9.936), ComMag (impact factor: 11.052), TMC (impact factor: 5.112), TNSE (impact factor: 5.213), JAIHC (impact factor: 4.594), AJSE (impact factor: 1.711), JNSM (impact factor: 2.25), and SIMPAT (impact factor: 2.219). Table 9 depicts the distribution of articles per publication channel.

TABLE 9
THE DISTRIBUTION OF THE STUDIES PER PUBLICATION CHANNEL

| Category | Publisher | Publication channel | Count |
|---|---|---|---|
| Journal | IEEE | IEEE Access | 3 |
| | | China Communications (ChinaComm) | 2 |
| | | IEEE Transactions on Industrial Informatics (TII) | 1 |
| | | IEEE Internet of Things Journal (IoT-J) | 1 |
| | | IEEE Communications Magazine (ComMag) | 1 |
| | | IEEE Transactions on Mobile Computing (TMC) | 1 |
| | | IEEE Transactions on Network Science and Engineering (TNSE) | 1 |
| | Springer | Journal of Ambient Intelligence and Humanized Computing (JAIHC) | 2 |
| | | Arabian Journal for Science and Engineering (AJSE) | 1 |
| | | Journal of Network and Systems Management (JNSM) | 1 |
| | Science Direct | Simulation Modelling Practice and Theory (SIMPAT) | 1 |
| Conference | IEEE | IEEE Global Communications Conference (GLOBECOM) | 2 |
| | | IEEE International Conference on Communications (ICC) | 1 |
| | | International Conference on Software, Telecommunications and Computer Networks (SoftCOM) | 1 |
| | | IEEE Vehicular Technology Conference (VTC Spring) | 1 |
| | | IEEE Symposium on Computers and Communications (ISCC) | 1 |
| | | IEEE International Conference on Ubiquitous Computing and Communications (IUCC) | 1 |
| | | International Conference on Big Data Security on Cloud (BigDataSecurity) | 1 |
| | | International Conference on Computing for Sustainable Global Development (INDIACom) | 1 |
| | | IEEE International Conference on Systems, Man, and Cybernetics (SMC) | 1 |
| | | IEEE International Conference on Computer and Information Technology (CIT). | 1 |
| | Springer | International Conference on Broadband and Wireless Computing, Communication and Applications (BWCCS) | 3 |
| | | International Conference on Intelligent Networking and Collaborative Systems (INCoS) | 2 |
| | | International Conference on P2P, Parallel, Grid, Cloud and Internet Computing (3PGCIC) | 2 |
| | | International Conference on Network-Based Information Systems (NBiS) | 2 |
| | | International Conference on Communications and Networking in China (EAI ChinaCom) | 1 |

### 5.1.3 Active Research Communities

After selecting the studies, we looked at the authors' affiliations. As shown in Table 10, only active communities that presented at least two articles, along with their primary research focuses, are listed. A remarkable number of articles that are associated with approximate methods have been published in the COMSATS University in Pakistan, The University of Sydney in Australia, the Sapienza University of Rome in Italy, and the Xidian University in China. Further, The University of Sydney in Australia and the Sapienza University of Rome in Italy studied exact methods. Besides, researchers in the Federal Rural University of Pernambuco in Brazil and the COMSATS University in Pakistan have focused on fundamental methods. Also, in the Mansoura University in Egypt and the COMSATS University in Pakistan, researchers concentrated more on hybrid methods.

TABLE 10
ACTIVE COMMUNITIES ON LOAD BALANCING MECHANISMS IN FOG COMPUTING

| Affiliation | Study ID | Research Focus |
| --- | --- | --- |
| COMSATS University, Pakistan | [37], [38], [30] [61], [62], [63] [68], [69], [70], [71] | Approximate, Fundamental, and Hybrid methods |
| The University of Sydney, Australia | [48], [51] | Approximate and Exact methods |
| Sapienza University of Rome, Italy | [40], [54] | Approximate and Exact methods |
| Xidian University, China | [43], [45] | Approximate methods |
| Federal Rural University of Pernambuco, Brazil | [65], [66] | Fundamental methods |
| Mansoura University, Egypt | [72], [73] | Hybrid methods |

## 5.2 Research Objectives, Techniques, and Evaluation Parameters

The review process of the chosen articles in load balancing mechanisms in fog computing was discussed in Section 4 in four main categories, including approximate, exact, fundamental, and hybrid methods. In Table 11, based on the obtained information from Tables 5, 6, 7, and 8, the main merits and demerits of these four categories are concisely proposed. According to Table 11, in the approximate methods category, the parameters such as security, reliability, and resource utilization are neglected. On the other hand, metrics such as latency, energy, and performance are more regarded. In exact methods category, better energy, cost, and response time are remarked, but it suffers from inadequate security, throughput, and reliability. In addition, the fundamental group has a better focus on energy, latency, performance, and cost. However, it suffers from inadequate throughput, security, security, and resource utilization. Finally, in the hybrid methods category, the parameters such as latency, cost, and energy are neglected. On the other hand, metrics such as reliability, security, and throughput are more regarded.

Moreover, now, the analytical and statistical reports of the research questions are presented based on the plan in Section 3.1 as follows:

- RQ 2: Which kind of classification in research approaches can be applied in fog systems load balancing?

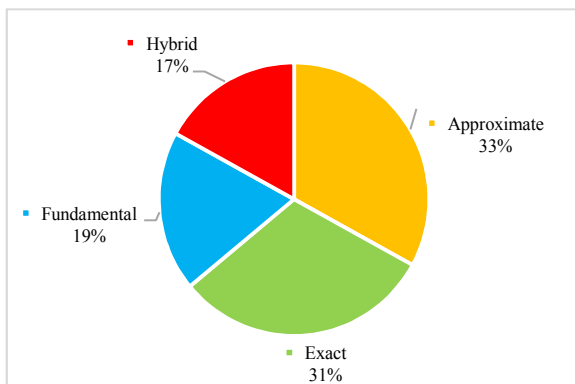

Fig. 6. The percentage of load balancing mechanisms in fog computing

The used methods in the load balancing of fog computing fall into four classes, and their statistic percentages are shown in Fig. 6. It was seen that the highest percentage of studies is conducted in approximate methods with 33%, while exact methods have 31%, fundamental methods have 19%, and hybrid methods have 17% of all types of used methods.

- RQ 3: What evaluation metrics are applied in load balancing mechanisms of fog computing?

In order to answer RQ 3 and to consider the literature, researchers have used various evaluation metrics. After applying (1) that obtains the percentage of each specific factor regarding the other mentioned factors and their percentages are illustrated in Figures 7 and 8. Figure 7 displays the percentage of evaluation metrics in load balancing mechanisms of fog computing with 27% for response time evaluation metric, with the highest percentage, and energy and cost come next with 17% each, processing time with 14%, resource utilization and scalability with 8% each, security with 4%, reliability with 3%, and finally, throughput and availability each have 1% of all evaluation metrics used in articles. As defined in Fig. 8, in approximate methods, 31% of the articles have made efforts to improve the response time, and 16% of articles have improved all factors of energy, scalability, and processing time. In exact methods, 28% of the studies have pushed for decreasing the response time, and 20% of the articles have tried to improve energy and cost each. In fundamental methods, 26% of studies have made attempts to improve response time, and 22% of them tried to decrease cost. Finally, in hybrid methods, 22% of the articles have made efforts to decrease the response time, and 19% of articles have improved cost.

$$Eval\_metric(i) = \frac{metric(i)}{\sum_{j=1}^{metric\_no} metric(j)} * 100 \qquad (1)$$

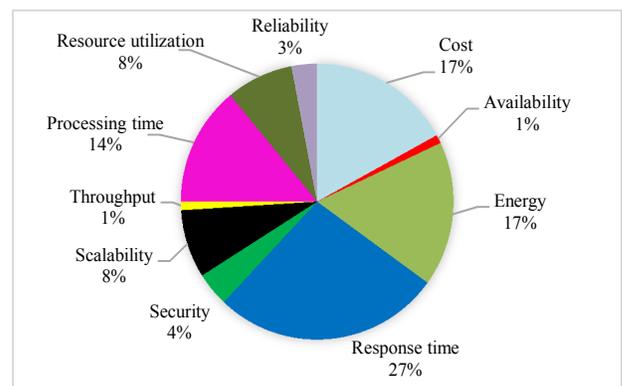

Fig. 7. The percentage of evaluation metrics in load balancing mechanisms of fog computing

- RQ 4: What popular evaluation tools are applied in load balancing mechanisms in fog computing?

Figure 9 defines the results provided by the comparisons in Tables 5, 6, 7, and 8, which illustrates the percentage of evaluation tools in the load balancing mechanisms in fog computing. Also, statistically, Fig. 9 shows the percentage of evaluation tools used for the review of the literature here in this article. The CloudAnalyst and MATLAB have 18% each, CloudSim has 7% of usage, NS-2/3, iFogSim, Java platform, JMeter, AMPL/CPLEX, Scyther, Mininet, work robots come next and custom

TABLE 11
MAIN MERITS AND DEMERITS OF APPROXIMATE, EXACT, FUNDAMENTAL, AND HYBRID METHODS

|  | Approximate | Exact | Fundamental | Hybrid |
|---|---|---|---|---|
| Merits | • Better latency<br>• Better energy<br>• Better performance | • Better energy<br>• Better response time<br>• Better cost | • Better energy<br>• Better latency<br>• Better performance<br>• Better cost | • Better latency<br>• Better cost<br>• Better energy |
| Demerits | • Unacceptable security<br>• Unacceptable reliability<br>• Unacceptable resource utilization | • Unacceptable security<br>• Unacceptable throughput<br>• Unacceptable reliability | • Unacceptable throughput<br>• Unacceptable security<br>• Unacceptable reliability<br>• Unacceptable resource utilization | • Unacceptable reliability<br>• Unacceptable security<br>• Unacceptable throughput |

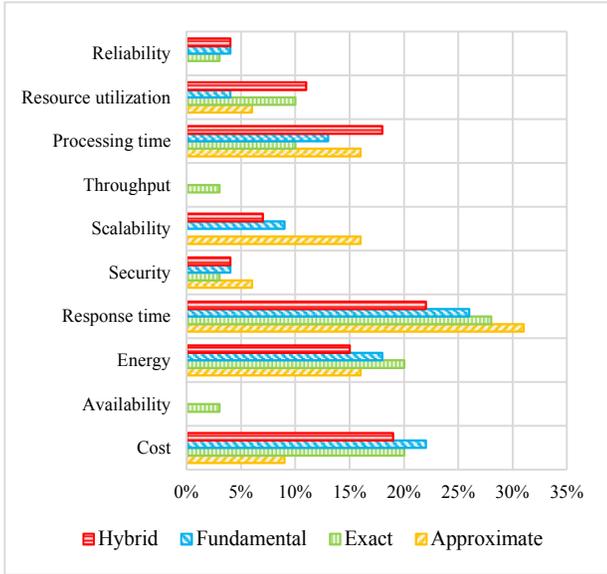

Fig. 8. The percentage of evaluation metrics in each classification.

simulator is the evaluation tools applied for these literature reviews.

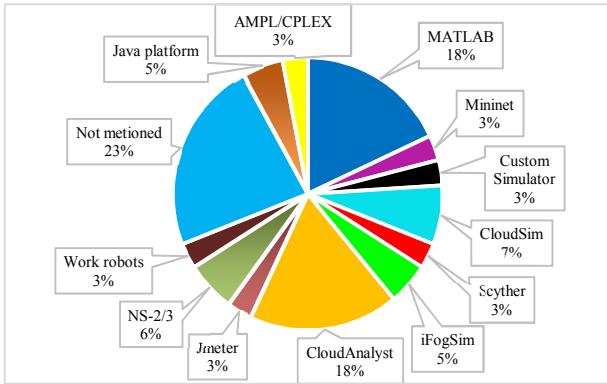

Fig. 9. The percentage of evaluation tools for load balancing mechanisms in fog computing

- RQ 5: What measurement techniques are used to assess the load balancing in fog computing?

On that basis of Tables 5, 6, 7, and 8, Fig. 10 displays the percentage of measurement environments in the load balancing mechanisms in fog computing. Simulation is the most used measurement environments with 85%, applied in four categories because actual environments have not yet been provided for fog computing; researchers mostly used simulation environments for measurement. Also, the formal verification used for measurement environments in some articles is in exact category. Finally, the prototype and example application applied for measurement environments in some studies are in the approximate and fundamental categories.

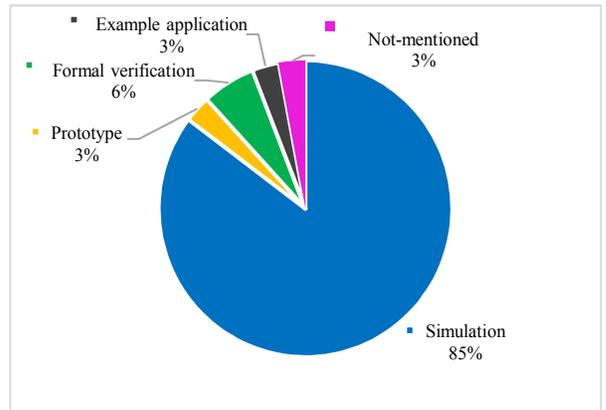

Fig. 10. The percentage of measurement environments in load balancing mechanisms of fog computing

## 6 OPEN ISSUES AND FUTURE DIRECTION

Based on the review, there are some key cases not having been investigated in the load balancing of fog network. So, in this section, the following answers to RQ 6 are presented.

- RQ 6: What are the open issues, future trends, and challenges of load balancing in fog computing?

◆ *Energy consumption and green fog*: In most of the techniques studied, the challenges of greenhouse gas and carbon emissions were not considered by researchers. In the topic of load balancing and fog computing, green fog computing and energy consumption can be important [74, 75]; However, only 17% of the reviewed articles focused on energy, which can optimize the popularity and efficacy of existing load balancing techniques. So, load balancing techniques, based on energy consumption and greenhouse gas and carbon emissions in fog environment, are up-and-coming and can be a direction for open issues.

◆ *Multi-objective optimization*: It is obvious that there is no mechanism, in particular, to define most of the QoS parameters to decide for load balancing in a fog network. Some of the mechanisms, for example, just regard energy, cost, or response time and ignore such parameters as scalability, reliability, security, etc. Therefore, multi-objective optimization in load balancing decision making needs to be expanded to bring some QoS parameters into consideration, and establishing tradeoff between different parameters might be a considerable open issue.

- *Optimal solutions*: In literature, most fog-based approaches to load balancing, such as scheduling, resource allocation, etc., are in terms of complexity in the NP-hard and NP-complete groups. Some meta-heuristic and heuristic algorithms have been used to solve them, but other optimization methods such as bacterial colony optimization [76], memetic [77], [78], [79], [80], Artificial Immune System [81], Imperialist Competitive Algorithm [82], grey wolf Optimizer [83], simulated annealing [84], firefly algorithm [85], lion optimizer algorithm [86], and glowworm swarm [87] are good directions for future works.
- *Implementation challenges*: Regarding that fog computing is under investigation, the real testbed is not yet available to most researchers, and it was found that 85% of the articles used simulator-based tools for their assessments. Because the results of cases like scheduling in the real environment can be different from the simulation environment, so it can be concluded that the implementation of the discussed mechanisms in the real testbed is really challenging.
- *Context-aware computing*: Context-aware computing assists load balancing in fog computing with new data to be applied for novel applications and to obtain knowledge establishment from the smart objects. What is more, the initial application of ontologies, as sources of learning, is recovered by context-aware computing that remains as a suspending case legally joining with the load balancing in fog computing smart objects [88, 89]. Using context-aware computing for developing load balancing techniques of fog network could be an exciting case for future trends.
- *SDN/NFV*: Considering the constraint of fog resources, the use of SDN in fog computing can provide load balancing more easily in fog network management. Applying network function virtualization (NFV) with fog computing can provide flexibility and speed in the management, construction, and deployment of novel applicant-based services. Using SDN/NFV [90, 91] technologies for supporting load balancing techniques in fog computing is an interesting case to be studied.
- *Scalability*: Some approaches in fog computing must be able to act on large scales. The validation of these approaches in small scales does not guarantee some nodes, devices, and related operations. Despite its importance, only 8% of the literature has focused on the scalability factor, and the articles related have been defined in small-scale scenarios, so it is an open challenge for future research.
- *Social networks analytics*: The expansion of social networks has created a good platform for their application in fog networks. The content transmitted through social networks generates big data. Social big data analytics can be applied to predict how fog resources are distributed. It can also predict resource and service requests in fog networks. Social networks such as Facebook, Twitter, and Instagram have created massive data and networks [92]. Therefore, applying social networks analytics to load balancing techniques of fog computing can be an attractive direction for future research.
- *Interoperability*: Because of the variety and distribution of fog nodes and sources, interoperability can be regarded as a key success element in the load balancing in IoT/fog domains. As consumers would not like to use just one service provider, they often look for their favorites and some important factors like cost, functionality, etc. Interoperability gives consumers the chance to move from one IoT/fog-based product to another, or to apply a combination of services and products to build smart environments, as they wish, in a customized way with load balancing [93]. So, regarding the interoperability as a key factor in combing the load balancing in IoT/fog-based services, it will be an interesting aspect for future studies.

## 7 LIMITATIONS

Because of systematic review-based methods, all existing studies might not have been evaluated. These types of review articles usually have limitations [5], but the results of the systematic review are mainly reliable [94]. The main limitations of this systematic review may include the following:

- Although several well-known databases, such as ScienceDirect, IEEE, ACM, Springer, etc. were used as reliable sources. However, it is not possible to guarantee that all the relevant studies are selected. So, some articles were neglected due to the mechanism described in Section 3.2.
- We did not consider non-peer-reviewed and non-English articles, chapters of books, thesis, review articles, short articles, and editorial papers.
- This article was divided into four groups: approximate, exact, fundamental, and hybrid. However, other taxonomies could have been possible.
- We presented six research questions to find their answers by investigating related papers. Other scholars may define some other research questions.

## 8 CONCLUSION

In this study, qualitative and quantitative research was conducted based on an SLR method on load balancing approaches in fog computing. Applying 915 studies published recently, between 2013 and August 2020, the authors offered the SLR-based method in this literature by using the exploration query. In addition, 36 studies focusing on load balancing in fog computing were examined. According to RQ 2, the used algorithms in the load balancing mechanisms of fog computing are categorized into four groups, with the highest percentage of researches done in approximate methods with 33%, exact methods with 31%, fundamental methods with 19%, and hybrid methods with 17% of all types of used methods. According to RQ 3, statistically, the percentage of evaluation metrics presented that the response time metric has the highest application in the assessment of the load balancing approaches by 27%, and cost and energy come next with 17% each. Based on RQ 4, for evaluation tools, it was observed that CloudAnalyst and MATLAB have the highest percentage of applying the simulation environment of case studies in load balancing approaches with 18% each. Also, with respect to the RQ 5, it was observed that 85% of studies applied a simulation environment for measuring the suggested load balancing approaches in fog computing because actual environments have not yet been provided for fog computing. Finally, based on RQ 6, the important open challenges and future directions of load balancing mechanisms in fog computing were described in detail.